\documentclass{nature}
\usepackage{graphicx}
\usepackage{dcolumn}
\usepackage{bm}
\usepackage{ifpdf}

\title{Unconditional Room Temperature Quantum Memory}

\author {M. Hosseini, G. Campbell, B. M. Sparkes, P. K. Lam and B. C. Buchler}

\begin{document}
\maketitle

\begin{affiliations}
 \item Centre for Quantum Computation and Communication Technology, Department of Quantum Science, Research School of Physics and Engineering, The Australian National University, Canberra, ACT 0200 Australia
\end{affiliations}


\begin{abstract}
Just as classical information systems require buffers and memory, the same is true for quantum information systems.  The potential that optical quantum information processing holds for revolutionising computation and communication is therefore driving significant research into developing optical quantum memory. A practical optical quantum memory must be able to store and recall quantum states on demand with high efficiency and low noise. Ideally, the platform for the memory would also be simple and inexpensive.  Here, we present a complete tomographic reconstruction of quantum states that have been stored in the ground states of rubidium in a vapour cell operating at around 80$^\circ$C. Without conditional measurements, we show recall fidelity up to 98\% for coherent pulses containing around one photon. In order to unambiguously verify that our memory beats the quantum no-cloning limit we employ state independent verification using conditional variance and signal transfer coefficients.
\end{abstract}

  \maketitle
While there are many physical systems for quantum computing \cite{Niskanen:SCqbit:science:2007,Petta:science:QD:2007,Home:ion:science:2009}, photons are an obvious choice for carrying quantum information over long distances due to their speed and weakly-interacting nature. A quantum memory that could perform noiseless and efficient storage of photonic quantum information is a cornerstone of quantum communication networks. Quantum memory is likely to be a necessary element of future optical quantum cryptographic networks \cite{Gisin:RevModPhys:QC:2002} and quantum logic gate operations \cite{Monroe:PRL:Qgate:1995}.

The natural bound that a quantum memory must overcome is the \textit{classical limit} \cite{Grosshans:2001p9578}. This is the storage performance that would be achieved via independent measurements of conjugate quantum observables.  Attempts to simultaneously measure conjugate variables always result in quantum back-action, so this measurement based approach to storage can never allow perfect reconstruction of the input state.  To unconditionally beat this limit, a quantum memory must have an efficiency greater than 50\% and work in a way that does not involve any projective measurement in order to avoid quantum back-action.  If this can be done, a new interesting performance benchmark can be surpassed, namely the \textit{no-cloning limit} \cite{Grosshans:2001p9578}. If this limit is beaten then it is guaranteed that the output of the quantum memory is the best possible copy of the original input state. For coherent states, the cloning fidelity limit is 0.68 \cite{Cerf:NGclon:PRL:2005}.

To date there have been many impressive demonstrations of quantum state storage in various systems including cold atomic ensembles \cite{BEC_1s,Jin:QM:sph,kimble_nature2008,Kimble4QM} and rare-earth ions in solid state \cite{Hedges:2010p11910,Saglamyurek:2011p12254,Clausen:2011p12253}.  In the present work we use an ensemble of Rb atoms in gaseous state \cite{lukin-QM-Rb-EIT, Julsgaard:2004p7617,Alex:PRL:2008:100,Honda:2008p4680} and store quantum states of light in the atomic ground states. Not only does our scheme exhibit high efficiency and high fidelity storage of quantum states, it does so in a common vapour cell operating above room temperature. We use a cell of $^{87}$Rb mixed with 0.5 Torr Kr buffer gas at $T\cong80^o$C. We verify storage of states beyond the no-cloning limit using quantum state tomography with near-coherent input states of photon numbers ranging from $n\approx$ 1 to 23.   Without the use of tuneable conditional measurement \cite{Eisaman:Nat:2005,Jin:QM:sph,Saglamyurek:2011p12254,Clausen:2011p12253}, we observed fidelity as high as 98\%.\\
\newline
\textbf{Scheme}\\

Our method stores optical information in the long-lived coherence between hyperfine ground states of warm Rb atoms. The Raman transition between these ground states is inhomogeneously broadened by a linearly varying Zeeman shift resulting from a magnetic field applied along the light propagation axis. A strong control field couples weak signal field to the broadened Raman transition, allowing the quantum state of the signal to be mapped onto the ground state atomic coherences with more than 99\% absorption efficiency. The stored information is retrieved as a photon echo by switching the sign of the magnetic gradient field, which has the effect of time-reversing the absorption process. The underlying photon-echo protocol has been previously described as longitudinal controlled reversible inhomogenous broadening (L-CRIB) \cite{Moiseev:CRIB:PRL:2001,Longdell:2008p8530,Alexander:JOS:2007,Nunn:PRL:2008} and gradient echo memory (GEM) \cite{Hetet:2008p7696,Hetet:2008p1839}. The three-level scheme in the present work is known as  Lambda Gradient Echo Memory ($\Lambda-$GEM) \cite{Hetet:2008p5840}. Earlier work has shown that this method is highly efficient, with demonstrated recall efficiency as high as 87\% \cite{Hosseini:natcomm:10}, while also allowing for spectral manipulation and recall of information in arbitrary order \cite{Hosseini:2009p8466,Buchler:2010p9599} (see the Supporting Materials for further details).

A two-level GEM  protocol \cite{Hetet:2008p1839,Hetet:2008p7696} has previously been applied to  a cryogenically cooled solid state sample where noiseless storage of optical pulses was demonstrated \cite{Hedges:2010p11910}. Compared to the two-level scheme, $\Lambda-$GEM is more versatile in terms of data manipulation and, in principle, is capable of long memory times due to the negligible coupling between the atomic ground states used for the storage. The price we pay for this versatility is the need for a control field.  This field is strong and, in some regimes, could result in the addition of noise to the stored quantum states due to the potential for non-linear interactions within the atomic vapour.  One of the key findings in the present work is that we can find a regime where $\Lambda$-GEM provides highly efficient storage without any memory contamination due to the control field.

  The signal and control fields have the same circular polarisation and are combined before the memory using a locked ring cavity. After the memory, the control field is filtered out using a gas cell containing $^{85}$Rb atoms \cite{Eisaman:Nat:2005,Jin:QM:sph}. The filtering provides $\sim$ 60 dB suppression of the control field with a signal field attenuation of only 1.5 dB.\\
 \newline
\textbf{Tomographic reconstruction}\\
To measure the noise performance of the memory and accomplish quantum state tomography we recorded more than 100,000 homodyne measurements for each input and output state. The input pulses had a duration of 2 $\mu s$ and were stored for 3 $\mu s$.  The bandwidth of the pulses was matched to the chosen memory bandwidth of 0.5 MHz to maximise the single-mode efficiency of the system.  The bandwidth of the GEM scheme is easily tuneable since it is determined by the applied magnetic broadening, although there is always a trade-off normalised between the efficiency and bandwidth when limited by the optical depth. The multimode capacity of our protocol has been demonstrated elsewhere\cite{Hosseini:natcomm:10}.

The control field was switched off for 1 $\mu s$ during storage to minimise decoherence due to scattering \cite{Hosseini:natcomm:10}. The storage time was made sufficiently long to avoid electronic noise associated with the magnetic field switching. To find the efficiency of our memory we compared stored pulses with reference pulses of identical size that were tuned far away from the atomic resonance so that they were fully transmitted by the memory gas cell. This was done for a range of different pulse amplitudes. Both the reference pulse and the stored pulse are thus subject to common transmission losses.  Our measured efficiency of 78$\pm5$\% therefore quantifies the memory process alone.  The other efficiency parameters of the experiment will be discussed in greater detail when we consider the the quantum nature of the memory. 

\begin{figure*}
     \includegraphics[width=15cm]{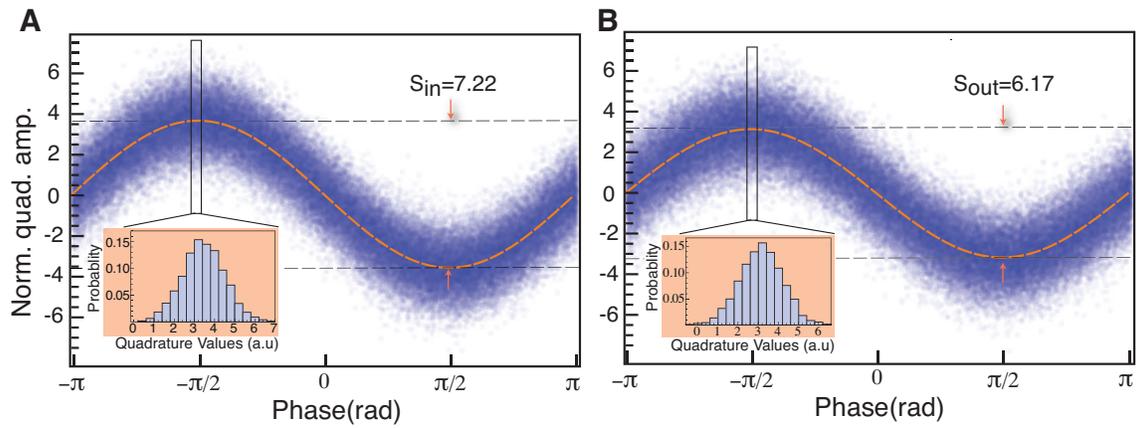}
       \centering
   \caption{ Normalised quadrature amplitude as a function of local oscillator phase for  \textbf{A}, input, and  \textbf{B}, echo pulses, normalised to the vacuum. The input pulse had a mean photon number of $\langle N \rangle = 3.4$. The amplitudes of input and output signals are shown as $S_{in}$ and $S_{out}$, respectively.  Insets show histograms of the quadrature values at the indicated phase. The plots each show 100,000 pulse quadrature measurements.}
  \label{quad}
\end{figure*}

To determine the phase of each pulse, we sent a strong reference pulse at a different frequency 9 $\mu s$ prior to the input pulse. This separation is small compared to the time scale of phase fluctuations in the experiment, so that we can reliably infer the phase of the input and echo pulses relative to our reference pulse. The error obtained from the least-squares fit to the pulse data indicates that phase estimation uncertainty is 29 mrad.  We integrated the amplitude of the input and the echo pulses over the pulse duration to find a quadrature value and then and used the reference pulse to associate a phase with each integrated quadrature value. Fig.~\ref{quad} A and B show the quadrature measurement results as a function of local oscillator phase for input and output pulses, respectively, with a mean input photon number of 3.4.

\begin{figure*}
     \includegraphics[width=14cm]{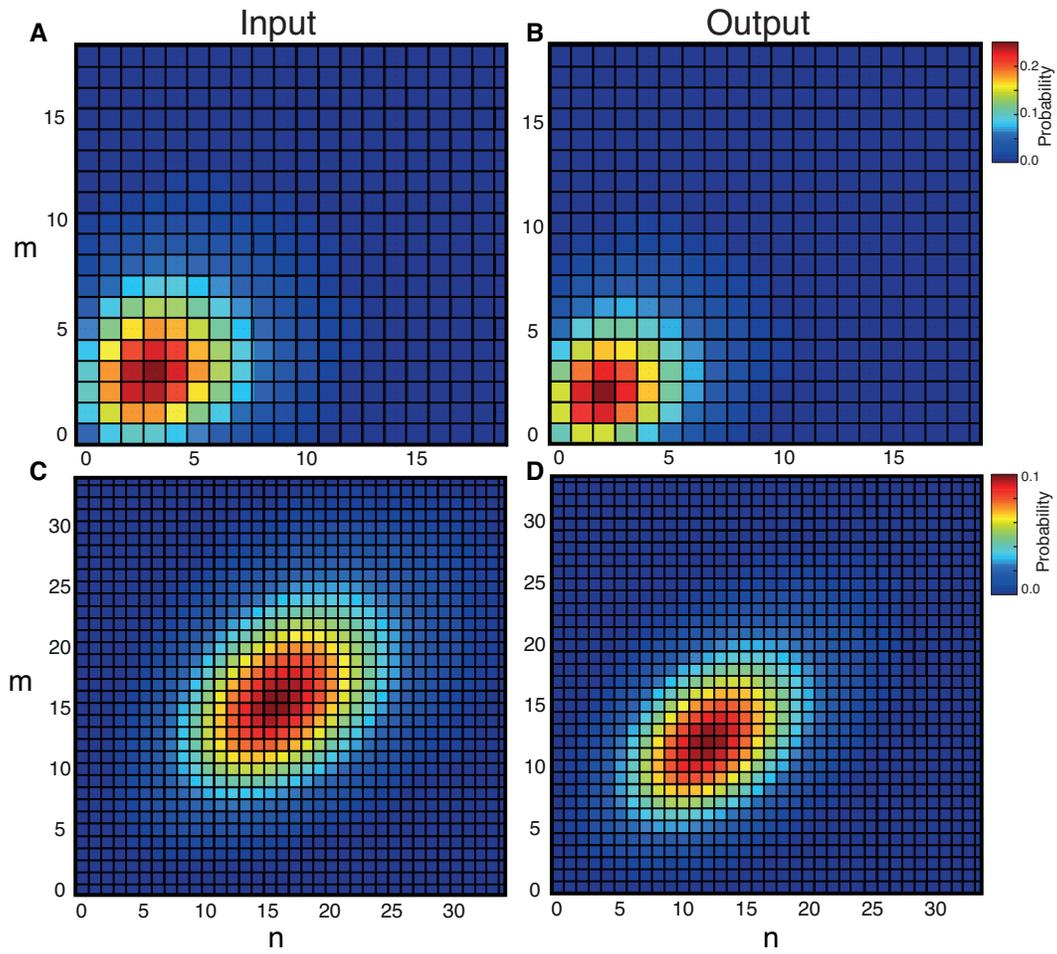}
       \centering
   \caption{ Density matrix elements for two sets of input and output pulses.  \textbf{A} and \textbf{B} Density matrix elements for input and output states, respectively, with $\langle N\rangle  = 3.4$ yielding a fidelity of 93\%. \textbf{C} and \textbf{D} Density matrix elements for input and output states, respectively,  with $\langle N\rangle =16$ yielding a fidelity of 82\%}
  \label{Rho}
 \end{figure*}

The quadrature measurements were used to reconstruct the density matrix elements, the results of which are plotted in Fig. \ref{Rho} for two coherent states with different amplitudes. This was done using an iterative maximum-likelihood reconstruction method  \cite{Maxlike} applied to the data collected from 100,000 pulses.

\begin{figure*}
     \includegraphics[width=14cm]{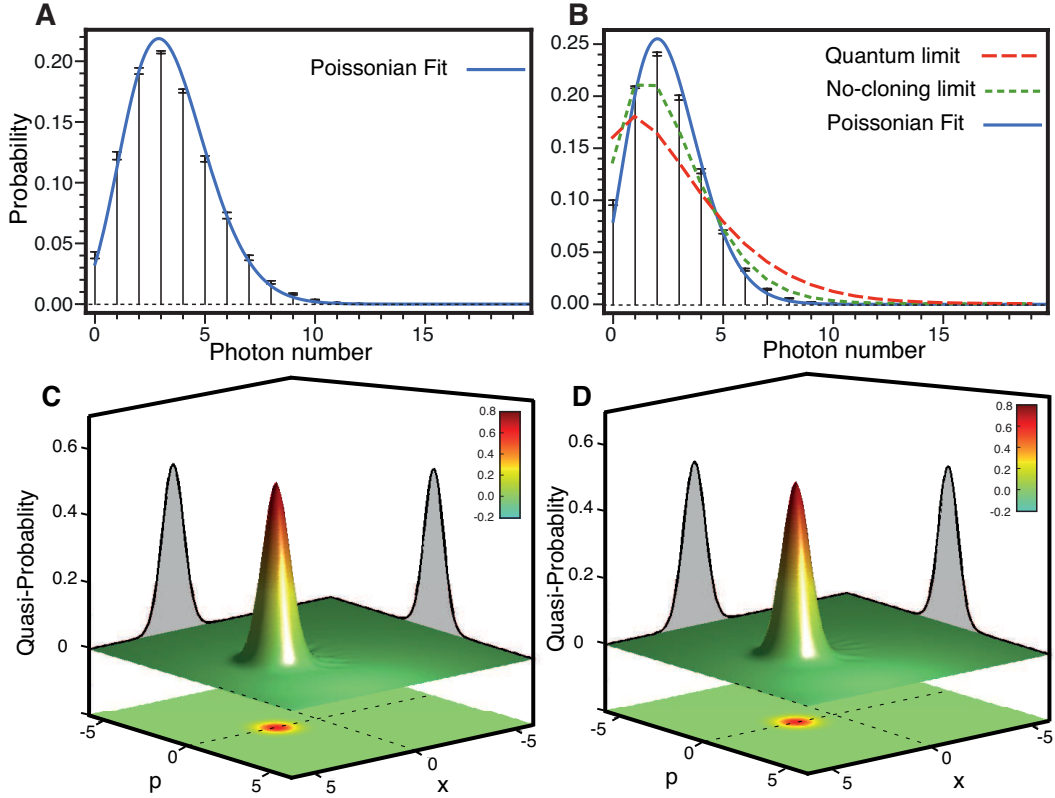}
       \centering
   \caption{ \textbf{A} and \textbf{B} Photon number distribution for input and output pulses, respectively. The blue solid lines show the fitted Poissonian distribution. The green dotted line represents the no-cloning limit and the dashed red line shows the boundary for the quantum limit. The error bars are statistical errors obtained from 100 subsets of data. \textbf{C} and \textbf{D} Reconstructed Wigner functions of input and output states for $\langle N\rangle = 3.4$. x and p represent the amplitude and phase of the coherent state, respectively.}
  \label{Phdis_Wig}
 \end{figure*}

The density matrix results allowed us to investigate the photon statistics of our light pulses before and after the memory. In Fig. \ref{Phdis_Wig} A and B, we plot the photon number distribution of the input and output pulses. These distributions are obtained  from the diagonal density matrix elements. The solid blue lines show a Poissonian distribution, fitted to the measured mean photon number of  3.4.  The good agreement of our data with this model shows that our output states are also near Poissonian, as we would expect for near coherent input states. This distribution can be compared to the photon statistics that would be obtained in the case of a memory with equal efficiency but contaminated by extra noise. To do this we assume equal amounts of Gaussian noise are added to the phase and amplitude quadratures of our output state and then find the resulting photon number distributions.  In Fig.~\ref{Phdis_Wig} B we show curves that illustrate the photon statistics we would obtain assuming Gaussian noise that degrades the fidelity of our memory to the classical and no-cloning limits. This data clearly shows that our memory does not introduce significant noise to the output pulses and easily exceed the no-cloning limit.\\
To get an intuitive picture for quantum-state tomography we reconstruct the Wigner function \cite{Walls:Milburn:QO}, which is a quasi-probability distribution in phase space. Among various phase space plots, the Wigner distribution is used frequently to measure probability in coordinate and momentum space.  Fig. \ref{Phdis_Wig} C and D show the reconstructed Wigner functions of the input and output states with $\langle N\rangle =3.4$. The projected probability distributions along the two marginal distributions, amplitude (x) and phase (p) represent a Gaussian distribution for x and p quadratures.\\

 \textbf{Fidelity and T-V characterisation}\\
In order to quantitatively characterise the memory performance in the quantum regime we analyse the storage fidelity by evaluating the overlap between the  input and output states. The fidelity  ($\mathcal{F}$)  can be computed as the overlap integral of the input and output Wigner functions or directly from the density matrix using the equation  $\mathcal{F}=|Tr(\sqrt{\sqrt{\rho_{in}}\rho_{out}\sqrt{\rho_{in}}})|^2$ \cite{Jeong:Fidelity:2004}. 

\begin{figure*}
     \includegraphics[width=14cm]{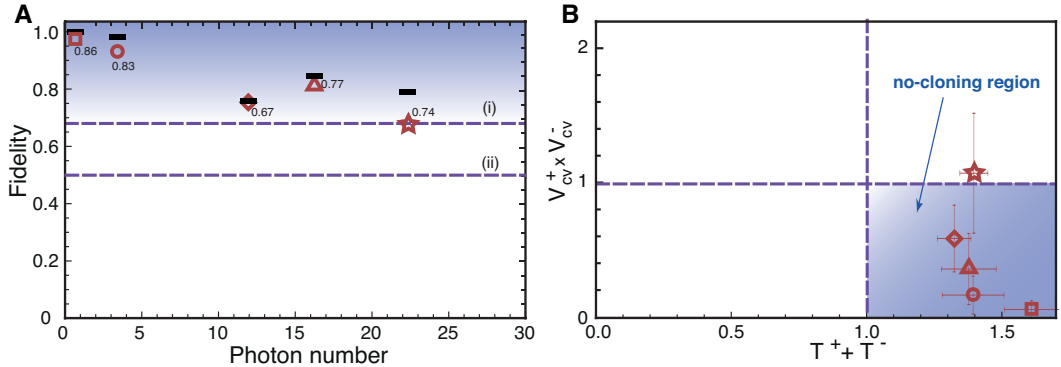}
       \centering
   \caption{ \textbf{A}: Fidelity of the memory for various optical states with different photon numbers. The measured memory efficiency for each data set is shown next to each symbol. Trace (i) shows the maximum fidelity that is expected from a classical memory. Trace (ii) is the no-cloning limit of 0.68. The small black lines show the maximum fidelity expected after a noiseless storage, taking into account the input noise and efficiency for each point. The statistical error in measuring the fidelity is smaller than the size of the symbols. \textbf{B}: The T-V diagram based on the same raw data as \textbf{A}. The symbols used here are the same as those in \textbf{A} to allow comparison of the data. The shaded region on the bottom-right is the no-cloning regime. The error bars represent the statistical uncertainties.}
  \label{TV}
 \end{figure*}

These results are presented in Fig. \ref{TV} A. The observed fidelity is as high as 93\% for $\langle N \rangle=3.4$ and 98\% for $\langle N \rangle=0.67$. In the limit of storing pulses with no photons, i.e. a vacuum, the efficiency of the memory no longer plays a role in determining the fidelity, since a memory with low efficiency can still output a pure vacuum state. For low photon numbers the fidelity is, however, sensitive to added noise. The high fidelity that we observe at low photon numbers is therefore indicative of a memory that does not add noise to the output state.   Also shown in this plot are the classical (trace(i)) and optimal fidelity (trace(ii)) limits for coherent states of 1/2 and 0.68, respectively.  All our data points are at or beyond the coherent state no-cloning limit.  This is, however, only of real significance for the two smallest photon numbers where the states are, to good approximation, coherent.  As expected from a real experiment, our pulses have some amount of noise above the vacuum fluctuations. This added noise, mostly due to small instabilities in our cavity locking servos, increases with photon number. Since the fidelity is highly state dependent the quantum and no-cloning benchmarks obtained for coherent state are not valid for states with higher photon numbers. Also shown on this figure are lines that indicate the maximum possible fidelity \cite{Ralph:JOSAB:2007} of our memory assuming the measured input state, the measured efficiency and the absence of any additional noise other than vacuum fluctuation introduced due to the sub-unity efficiency.  Except for the largest state ( $\langle N\rangle = 22.4$), our memory performs close to or at this limit for all of our data points, again indicating that our memory does not add any substantial noise to the stored quantum state.

 As the above analysis shows, the state dependent nature of the fidelity means that it is not an easy-to-use measure of the memory performance. In the case where the memory is being probed with various input states with different levels of added noise, each input state has its own unique no-cloning limit for fidelity. To unambiguously quantify the performance of our memory it would be advantageous to use a state independent criterion. This can be done using a signal-transfer and conditional-variance characterisation known as a T-V diagram. This method was originally proposed for characterising quantum non-demolition measurements \cite{Poizat:1994p12347} and later adapted to quantum teleportation \cite{Ralph:PKL:TV:1998,Bowen:2003p837} and quantum memory \cite{Hetet:2008p1841,Ortalo:2009p12255}.  The conditional variance of the amplitude $V^{+}$ and phase $V^{-}$ quadratures is a measure of the noise added by the memory.  An ideal memory adds no noise so the conditional variance between the input and output would be 0. The classical limit would be the case where the noise added by the memory is one unit of vacuum noise on each quadrature so that $V^{+}=V^{-}=1$.  The amplitude and phase signal transfer coefficients ($T^+$ and $T^-$ respectively) are a measure of how well the memory preserves a signal.  If the signal-to-noise ratio of the output is equal to the input, as would be the case for an ideal memory, then the transfer coefficient is unity.  The classical limit is $T^{+}=T^{-}=0.5$. It can be shown that if the two quantum benchmarks of $ V^+_{cv} \times V^{-}_{cv} \leq 1 $ and $T^ {+} + T ^{-} \geq 1$ are satisfied then the memory device surpasses the no-cloning limit  \cite{Grosshans:2001p9578}. As presented in Fig. \ref{TV} B, almost all of the experimental data points corresponding to various states are within the no-cloning region. 

When calculating the conditional variance it is important to account for the total detection efficiency of the experiment. In our analysis the quantum efficiency of the detectors (90\%), fringe visibility of the homodyne (97\%) and transmission of the signal through the filtering cell (70\%)  have been taken into account while calculating the conditional variances by extrapolating the variances of the input and output to the state prior to these losses. With this state independent measurement, the results demonstrate that our system has convincingly surpassed the no-cloning limit of quantum memory for a range of photon numbers. 

In the current experiment, the coherence time of the memory ($\sim $10 $\mu s$) is limited by the diffusion and collision of atoms. It was recently shown that by preparing cells with single-compound alkene-based coatings, spin relaxation times of up to few seconds can be easily achieved even at high temperatures \cite{Polzik_cell_coating}. This spin relaxation time is comparable to the best coherence time measured in cold atomic ensembles. In terms of the miniaturisation of these types of memories, extensive work has been done to manufacture microscopic vapour cells for alkali atoms \cite{micocell:natph,microRbcell:2008, onchipRb:nphot:2007,microcells:OL:2010}.
Hollow-core waveguides also show great promise in developing integrated coherent photonic structures\cite{Wu:nph2010chip}. All of these developments together with the results presented here suggest that Rb vapour could be a reliable and scalable platform for quantum memory.

Correspondence should be addressed to Ben Buchler, e-mail: ben.buchler@anu.edu.au

 \textbf{Acknowledgements}  The authors thank Julien Bernu for providing us with a Matlab code to perform maximum likelihood reconstruction and Gabriel H\'etet, Matthew Sellars and Tim Ralph for numerous enlightening discussions. This research was conducted by the Australian Research Council Centre of Excellence for Quantum Computation and Communication Technology (project number CE110001027).
 
 \textbf{Competing financial interests} The authors declare no competing financial interests.

\textbf{Contributions}
Experiments, measurements and data analysis were performed by M. H. with assistance of B. M. S. and G. C. for data collection and experimental preparation.  The project was planned and supervised by B.C.B. and P.K.L.  The manuscript was written by M. H and B.C.B. and G. C with the assistance of all other authors.

\bibliographystyle{naturemag}


\begin{thebibliography}{10}
\expandafter\ifx\csname url\endcsname\relax
  \def\url#1{\texttt{#1}}\fi
\expandafter\ifx\csname urlprefix\endcsname\relax\def\urlprefix{URL }\fi
\providecommand{\bibinfo}[2]{#2}
\providecommand{\eprint}[2][]{\url{#2}}

\bibitem{Niskanen:SCqbit:science:2007}
\bibinfo{author}{Niskanen, A.~O.} \emph{et~al.}
\newblock \bibinfo{title}{Quantum coherent tunable coupling of superconducting
  qubits}.
\newblock \emph{\bibinfo{journal}{Science}} \textbf{\bibinfo{volume}{316}},
  \bibinfo{pages}{723--726} (\bibinfo{year}{2007}).

\bibitem{Petta:science:QD:2007}
\bibinfo{author}{Petta, J.~R.} \& \bibinfo{author}{et~al.}
\newblock \bibinfo{title}{Coherent manipulation of coupled electron spins in
  semiconductor quantum dots}.
\newblock \emph{\bibinfo{journal}{Science}} \textbf{\bibinfo{volume}{309}},
  \bibinfo{pages}{2180--2184} (\bibinfo{year}{2005}).

\bibitem{Home:ion:science:2009}
\bibinfo{author}{Home, J.~P.} \& \bibinfo{author}{et~al.}
\newblock \bibinfo{title}{Complete methods set for scalable ion trap quantum
  information processing}.
\newblock \emph{\bibinfo{journal}{Science}} \textbf{\bibinfo{volume}{325}},
  \bibinfo{pages}{1227--1230} (\bibinfo{year}{2009}).

\bibitem{Gisin:RevModPhys:QC:2002}
\bibinfo{author}{Gisin, N.}, \bibinfo{author}{Ribordy, G.},
  \bibinfo{author}{Tittel, W.} \& \bibinfo{author}{Zbinden, H.}
\newblock \bibinfo{title}{Quantum cryptography}.
\newblock \emph{\bibinfo{journal}{Rev. of Mod. Phys.}}
  \textbf{\bibinfo{volume}{74}}, \bibinfo{pages}{145--195}
  (\bibinfo{year}{2002}).

\bibitem{Monroe:PRL:Qgate:1995}
\bibinfo{author}{Monroe, C.} \& \bibinfo{author}{et~al.}
\newblock \bibinfo{title}{Demonstration of a fundamental quantum logic gate}.
\newblock \emph{\bibinfo{journal}{Phys. Rev. Lett.}}
  \textbf{\bibinfo{volume}{75}}, \bibinfo{pages}{4714--4717}
  (\bibinfo{year}{1995}).

\bibitem{Grosshans:2001p9578}
\bibinfo{author}{Grosshans, F.} \& \bibinfo{author}{Grangier, P.}
\newblock \bibinfo{title}{Quantum cloning and teleportation criteria for
  continuous quantum variables}.
\newblock \emph{\bibinfo{journal}{Phys. Rev. A}} \textbf{\bibinfo{volume}{64}},
  \bibinfo{pages}{010301--4} (\bibinfo{year}{2001}).

\bibitem{Cerf:NGclon:PRL:2005}
\bibinfo{author}{Cerf, N.~J.}, \bibinfo{author}{Kruger, O.},
  \bibinfo{author}{Navez, P.}, \bibinfo{author}{Werner, R.~F.} \&
  \bibinfo{author}{Wolf, M.~M.}
\newblock \bibinfo{title}{Non-gaussian cloning of quantum coherent states is
  optimal}.
\newblock \emph{\bibinfo{journal}{Phys. Rev. Lett.}}
  \textbf{\bibinfo{volume}{95}}, \bibinfo{pages}{070501--4}
  (\bibinfo{year}{2005}).

\bibitem{BEC_1s}
\bibinfo{author}{Zhang, R.}, \bibinfo{author}{Garner, S.~R.} \&
  \bibinfo{author}{Hau, L.~V.}
\newblock \bibinfo{title}{Creation of long-term coherent optical memory via
  controlled nonlinear interactions in bose-einstein condensates}.
\newblock \emph{\bibinfo{journal}{Phys. Rev. Lett.}}
  \textbf{\bibinfo{volume}{103}}, \bibinfo{pages}{233602--4}
  (\bibinfo{year}{2009}).

\bibitem{Jin:QM:sph}
\bibinfo{author}{Jin, X.-M.} \& \bibinfo{author}{et~al.}
\newblock \bibinfo{title}{Quantum interface between frequency-uncorrelated
  down-converted entanglement and atomic-ensemble quantum memory}.
\newblock \emph{\bibinfo{journal}{arXiv:1004.4691v1 [quant-ph]}}
  (\bibinfo{year}{2010}).

\bibitem{kimble_nature2008}
\bibinfo{author}{Choi, K.~S.}, \bibinfo{author}{Deng, H.},
  \bibinfo{author}{Laurat, J.} \& \bibinfo{author}{Kimble, H.~J.}
\newblock \bibinfo{title}{Mapping photonic entanglement into and out of a
  quantum memory}.
\newblock \emph{\bibinfo{journal}{Nature}} \bibinfo{pages}{67--71}
  (\bibinfo{year}{2008}).

\bibitem{Kimble4QM}
\bibinfo{author}{Choi, K.~S.}, \bibinfo{author}{Goban, A.},
  \bibinfo{author}{Papp, S.~B.}, \bibinfo{author}{van Enk, S.~J.} \&
  \bibinfo{author}{Kimble, H.~J.}
\newblock \bibinfo{title}{Entanglement of spin waves among four quantum
  memories}.
\newblock \emph{\bibinfo{journal}{Nature}} \textbf{\bibinfo{volume}{468}},
  \bibinfo{pages}{412--416} (\bibinfo{year}{2010}).

\bibitem{Hedges:2010p11910}
\bibinfo{author}{Hedges, M.}, \bibinfo{author}{Longdell, J.},
  \bibinfo{author}{Li, Y.} \& \bibinfo{author}{Sellars, M.}
\newblock \bibinfo{title}{Efficient quantum memory for light}.
\newblock \emph{\bibinfo{journal}{Nature}} \textbf{\bibinfo{volume}{465}},
  \bibinfo{pages}{1052--1056} (\bibinfo{year}{2010}).

\bibitem{Saglamyurek:2011p12254}
\bibinfo{author}{Saglamyurek, E.} \emph{et~al.}
\newblock \bibinfo{title}{Broadband waveguide quantum memory for entangled
  photons}.
\newblock \emph{\bibinfo{journal}{Nature}} \textbf{\bibinfo{volume}{469}},
  \bibinfo{pages}{512--515} (\bibinfo{year}{2011}).

\bibitem{Clausen:2011p12253}
\bibinfo{author}{Clausen, C.} \emph{et~al.}
\newblock \bibinfo{title}{Quantum storage of photonic entanglement in a
  crystal}.
\newblock \emph{\bibinfo{journal}{Nature}} \textbf{\bibinfo{volume}{469}},
  \bibinfo{pages}{508--511} (\bibinfo{year}{2011}).

\bibitem{lukin-QM-Rb-EIT}
\bibinfo{author}{van~der Wal, C.~H.} \emph{et~al.}
\newblock \bibinfo{title}{Atomic memory for correlated photon states}.
\newblock \emph{\bibinfo{journal}{Science}} \textbf{\bibinfo{volume}{301}},
  \bibinfo{pages}{196--200} (\bibinfo{year}{2003}).

\bibitem{Julsgaard:2004p7617}
\bibinfo{author}{Julsgaard, B.}, \bibinfo{author}{Sherson, J.},
  \bibinfo{author}{Cirac, J.}, \bibinfo{author}{Fiurasek, J.} \&
  \bibinfo{author}{Polzik, E.}
\newblock \bibinfo{title}{Experimental demonstration of quantum memory for
  light}.
\newblock \emph{\bibinfo{journal}{Nature}} \textbf{\bibinfo{volume}{432}},
  \bibinfo{pages}{482--486} (\bibinfo{year}{2004}).

\bibitem{Alex:PRL:2008:100}
\bibinfo{author}{Appel, J.}, \bibinfo{author}{Figueroa, E.},
  \bibinfo{author}{Korystov, D.}, \bibinfo{author}{Lobino, M.} \&
  \bibinfo{author}{Lvovsky, A.~I.}
\newblock \bibinfo{title}{Quantum memory for squeezed light}.
\newblock \emph{\bibinfo{journal}{Phys. Rev. Lett.}}
  \textbf{\bibinfo{volume}{100}}, \bibinfo{pages}{093602--4}
  (\bibinfo{year}{2008}).

\bibitem{Honda:2008p4680}
\bibinfo{author}{Honda, K.} \emph{et~al.}
\newblock \bibinfo{title}{Storage and retrieval of a squeezed vacuum}.
\newblock \emph{\bibinfo{journal}{Phys. Rev. Lett.}}
  \textbf{\bibinfo{volume}{100}}, \bibinfo{pages}{093601--4}
  (\bibinfo{year}{2008}).

\bibitem{Eisaman:Nat:2005}
\bibinfo{author}{Eisaman, M.~D.} \& \bibinfo{author}{et~al.}
\newblock \bibinfo{title}{Electromagnetically induced transparency with tunable
  single-photon pulses}.
\newblock \emph{\bibinfo{journal}{Nature}} \textbf{\bibinfo{volume}{438}},
  \bibinfo{pages}{837--841} (\bibinfo{year}{2005}).

\bibitem{Moiseev:CRIB:PRL:2001}
\bibinfo{author}{Moiseev, S.~A.} \& \bibinfo{author}{Kroll, S.}
\newblock \bibinfo{title}{Complete reconstruction of the quantum state of a
  single-photon wave packet absorbed by a doppler-broadened transition}.
\newblock \emph{\bibinfo{journal}{Phys. Rev. Lett}}
  \textbf{\bibinfo{volume}{87}}, \bibinfo{pages}{173601--4}
  (\bibinfo{year}{2001}).

\bibitem{Longdell:2008p8530}
\bibinfo{author}{Longdell, J.~J.}, \bibinfo{author}{Hetet, G.},
  \bibinfo{author}{Lam, P.~K.} \& \bibinfo{author}{Sellars, M.~J.}
\newblock \bibinfo{title}{Analytic treatment of controlled reversible
  inhomogeneous broadening quantum memories for light using two-level atoms}.
\newblock \emph{\bibinfo{journal}{Phys. Rev. A}} \textbf{\bibinfo{volume}{78}},
  \bibinfo{pages}{032337} (\bibinfo{year}{2008}).

\bibitem{Alexander:JOS:2007}
\bibinfo{author}{Alexander, A.~L.}, \bibinfo{author}{Longdell, J.~J.} \&
  \bibinfo{author}{Sellars, M.~J.}
\newblock \bibinfo{title}{Measurement of the ground-state hyperfine coherence
  time of 151eu$^{3+}$:y$_2$sio$_5$}.
\newblock \emph{\bibinfo{journal}{J. Opt. Soc. Am. B}}
  \textbf{\bibinfo{volume}{24}}, \bibinfo{pages}{2479--2482}
  (\bibinfo{year}{2007}).

\bibitem{Nunn:PRL:2008}
\bibinfo{author}{Nunn, J.} \& \bibinfo{author}{et. al.}
\newblock \bibinfo{title}{Multimode memories in atomic ensembles}.
\newblock \emph{\bibinfo{journal}{Phys. Rev. Lett.}}
  \textbf{\bibinfo{volume}{101}}, \bibinfo{pages}{260502--4}
  (\bibinfo{year}{2008}).

\bibitem{Hetet:2008p7696}
\bibinfo{author}{Hetet, G.}, \bibinfo{author}{Longdell, J.~J.},
  \bibinfo{author}{Sellars, M.~J.}, \bibinfo{author}{Lam, P.~K.} \&
  \bibinfo{author}{Buchler, B.~C.}
\newblock \bibinfo{title}{Multimodal properties and dynamics of gradient echo
  quantum memory}.
\newblock \emph{\bibinfo{journal}{Phys. Rev. Lett.}}
  \textbf{\bibinfo{volume}{101}}, \bibinfo{pages}{203601--4}
  (\bibinfo{year}{2008}).

\bibitem{Hetet:2008p1839}
\bibinfo{author}{Hetet, G.}, \bibinfo{author}{Longdell, J.~J.},
  \bibinfo{author}{Alexander, A.~L.}, \bibinfo{author}{Lam, P.~K.} \&
  \bibinfo{author}{Sellars, M.~J.}
\newblock \bibinfo{title}{Electro-optic quantum memory for light using
  two-level atoms}.
\newblock \emph{\bibinfo{journal}{Phys. Rev. Lett.}}
  \textbf{\bibinfo{volume}{100}}, \bibinfo{pages}{023601--4}
  (\bibinfo{year}{2008}).

\bibitem{Hetet:2008p5840}
\bibinfo{author}{H{\'e}tet, G.} \emph{et~al.}
\newblock \bibinfo{title}{Photon echoes generated by reversing magnetic field
  gradients in a rubidium vapor}.
\newblock \emph{\bibinfo{journal}{Opt. Lett.}} \textbf{\bibinfo{volume}{33}},
  \bibinfo{pages}{2323--2325} (\bibinfo{year}{2008}).

\bibitem{Hosseini:natcomm:10}
\bibinfo{author}{Hosseini, M.}, \bibinfo{author}{Sparkes, B.~M.},
  \bibinfo{author}{Campbell, G.}, \bibinfo{author}{Buchler, B.~C.} \&
  \bibinfo{author}{Lam, P.~K.}
\newblock \bibinfo{title}{High efficiency coherent optical memory with warm
  rubidium vapour}.
\newblock \emph{\bibinfo{journal}{Nat. Commun.}} \textbf{\bibinfo{volume}{2}},
  \bibinfo{pages}{1--5} (\bibinfo{year}{2011}).

\bibitem{Hosseini:2009p8466}
\bibinfo{author}{Hosseini, M.} \emph{et~al.}
\newblock \bibinfo{title}{Coherent optical pulse sequencer for quantum
  applications}.
\newblock \emph{\bibinfo{journal}{Nature}} \textbf{\bibinfo{volume}{461}},
  \bibinfo{pages}{241--245} (\bibinfo{year}{2009}).

\bibitem{Buchler:2010p9599}
\bibinfo{author}{Buchler, B.~C.}, \bibinfo{author}{Hosseini, M.},
  \bibinfo{author}{H{\'e}tet, G.}, \bibinfo{author}{Sparkes, B.~M.} \&
  \bibinfo{author}{Lam, P.~K.}
\newblock \bibinfo{title}{Precision spectral manipulation of optical pulses
  using a coherent photon echo memory}.
\newblock \emph{\bibinfo{journal}{Opt. Lett.}} \textbf{\bibinfo{volume}{35}},
  \bibinfo{pages}{1091--1093} (\bibinfo{year}{2010}).

\bibitem{Maxlike}
\bibinfo{author}{Reh{\`a}cek, J.}, \bibinfo{author}{Hradil, Z.} \&
  \bibinfo{author}{Jezek, M.}
\newblock \bibinfo{title}{Iterative algorithm for reconstruction of entangled
  states}.
\newblock \emph{\bibinfo{journal}{Phys. Rev. A}} \textbf{\bibinfo{volume}{63}},
  \bibinfo{pages}{040303--4} (\bibinfo{year}{2002}).

\bibitem{Walls:Milburn:QO}
\bibinfo{author}{Walls, D.} \& \bibinfo{author}{Milburn, G.}
\newblock \emph{\bibinfo{title}{Quantum Optics}} (\bibinfo{publisher}{Springer,
  Berlin}, \bibinfo{year}{1994}).

\bibitem{Jeong:Fidelity:2004}
\bibinfo{author}{Jeong, H.}, \bibinfo{author}{Ralph, T.~C.} \&
  \bibinfo{author}{Bowen, W.~P.}
\newblock \bibinfo{title}{Quantum and classical fidelity for guassian states}.
\newblock \emph{\bibinfo{journal}{J of Optical Society of America B- Opt.
  Phys.}} \textbf{\bibinfo{volume}{24}}, \bibinfo{pages}{355--362}
  (\bibinfo{year}{2007}).

\bibitem{Ralph:JOSAB:2007}
\bibinfo{author}{Jeong, H.}, \bibinfo{author}{Ralph, T.~C.} \&
  \bibinfo{author}{Bowen, W.~P.}
\newblock \bibinfo{title}{Quantum and classical fidelities for gaussian
  states}.
\newblock \emph{\bibinfo{journal}{J. Opt. Soc. Am. B}}
  \textbf{\bibinfo{volume}{24}}, \bibinfo{pages}{355--362}
  (\bibinfo{year}{2007}).

\bibitem{Poizat:1994p12347}
\bibinfo{author}{Poizat, J.-P.}, \bibinfo{author}{Roch, J.-F.} \&
  \bibinfo{author}{Grangier, P.}
\newblock \bibinfo{title}{Characterization of quantum non-demolition
  measurements in optics}.
\newblock \emph{\bibinfo{journal}{Ann. Phys. Fr.}}
  \textbf{\bibinfo{volume}{19}}, \bibinfo{pages}{265--297}
  (\bibinfo{year}{1994}).

\bibitem{Ralph:PKL:TV:1998}
\bibinfo{author}{Ralph, T.~C.} \& \bibinfo{author}{Lam, P.~K.}
\newblock \bibinfo{title}{Teleportation with bright squeezed light}.
\newblock \emph{\bibinfo{journal}{Phys. Rev. Lett.}}
  \textbf{\bibinfo{volume}{81}}, \bibinfo{pages}{5668--4}
  (\bibinfo{year}{1998}).

\bibitem{Bowen:2003p837}
\bibinfo{author}{Bowen, W.} \emph{et~al.}
\newblock \bibinfo{title}{Experimental investigation of continuous-variable
  quantum teleportation}.
\newblock \emph{\bibinfo{journal}{Phys. Rev. A}} \textbf{\bibinfo{volume}{67}},
  \bibinfo{pages}{032302--4} (\bibinfo{year}{2003}).

\bibitem{Hetet:2008p1841}
\bibinfo{author}{Hetet, G.}, \bibinfo{author}{Peng, A.},
  \bibinfo{author}{Johnsson, M.~T.}, \bibinfo{author}{Hope, J.~J.} \&
  \bibinfo{author}{Lam, P.~K.}
\newblock \bibinfo{title}{Characterization of
  electromagnetically-induced-transparency-based continuous-variable quantum
  memories}.
\newblock \emph{\bibinfo{journal}{Phys. Rev. A}} \textbf{\bibinfo{volume}{77}},
  \bibinfo{pages}{012323--16} (\bibinfo{year}{2008}).

\bibitem{Ortalo:2009p12255}
\bibinfo{author}{Ortalo, J.} \emph{et~al.}
\newblock \bibinfo{title}{Atomic-ensemble-based quantum memory for sideband
  modulations}.
\newblock \emph{\bibinfo{journal}{Journal of Physics B: Atomic, Molecular and
  Optical Physics}} \textbf{\bibinfo{volume}{42}}, \bibinfo{pages}{114010--8}
  (\bibinfo{year}{2009}).

\bibitem{Polzik_cell_coating}
\bibinfo{author}{Balabas, M.~V.} \emph{et~al.}
\newblock \bibinfo{title}{High quality anti-relaxation coating material for
  alkali atom vapor cells}.
\newblock \emph{\bibinfo{journal}{Opt. Exp.}} \textbf{\bibinfo{volume}{18}},
  \bibinfo{pages}{5825--5830} (\bibinfo{year}{2010}).

\bibitem{micocell:natph}
\bibinfo{author}{Kubler, H.}, \bibinfo{author}{Shaffer, J.~P.},
  \bibinfo{author}{Baluktsian, T.}, \bibinfo{author}{Low, R.} \&
  \bibinfo{author}{Pfau1, T.}
\newblock \bibinfo{title}{Coherent excitation of rydberg atoms in
  micrometre-sized atomic vapour cells}.
\newblock \emph{\bibinfo{journal}{Nature Photonics}}
  \textbf{\bibinfo{volume}{4}}, \bibinfo{pages}{112--116}
  (\bibinfo{year}{2010}).

\bibitem{microRbcell:2008}
\bibinfo{author}{Eklunda, E.~J.}, \bibinfo{author}{Shkela, A.~M.},
  \bibinfo{author}{Knappeb, S.}, \bibinfo{author}{Donleyb, E.} \&
  \bibinfo{author}{Kitching, J.}
\newblock
  \bibinfo{title}{Glass-blownsphericalmicrocellsforchip-scaleatomicdevices}.
\newblock \emph{\bibinfo{journal}{Sensors and Actuators A}}
  \textbf{\bibinfo{volume}{143}}, \bibinfo{pages}{175--180}
  (\bibinfo{year}{2008}).

\bibitem{onchipRb:nphot:2007}
\bibinfo{author}{Yang, W.} \& \bibinfo{author}{et~al.}
\newblock \bibinfo{title}{Atomic spectroscopy on a chip}.
\newblock \emph{\bibinfo{journal}{Nature Phot.}} \textbf{\bibinfo{volume}{1}},
  \bibinfo{pages}{331--335} (\bibinfo{year}{2007}).

\bibitem{microcells:OL:2010}
\bibinfo{author}{Baluktsian, T.} \& \bibinfo{author}{et~al.}
\newblock \bibinfo{title}{Fabrication method for microscopic vapor cells for
  alkali aroms}.
\newblock \emph{\bibinfo{journal}{Opt. Lett.}} \textbf{\bibinfo{volume}{35}},
  \bibinfo{pages}{1950--1952} (\bibinfo{year}{2010}).

\bibitem{Wu:nph2010chip}
\bibinfo{author}{Wu, B.} \emph{et~al.}
\newblock \bibinfo{title}{Slow light on a chip via atomic quantum state
  control}.
\newblock \emph{\bibinfo{journal}{Nature Photonics}}
  \textbf{\bibinfo{volume}{4}}, \bibinfo{pages}{776--779}
  (\bibinfo{year}{2010}).

\end{thebibliography}

\clearpage

\end{document}